\documentclass[structabstract]{aa} 
\usepackage{graphicx}
\usepackage{lipsum}
\usepackage{subcaption}        
\usepackage{lscape}    
\usepackage{placeins}   
\usepackage[T1]{fontenc}

\usepackage[varg]{txfonts}
\usepackage{amsmath}	
\usepackage{amssymb}	
\usepackage{newtxtext,newtxmath}
\usepackage[normalem]{ulem}

\usepackage{natbib}
\usepackage[hidelinks]{hyperref}
\hypersetup{
     colorlinks   = true,
     linkcolor    = blue,
     citecolor    = blue
}
\newcommand\be{\begin{equation}}
\newcommand\ee{\end{equation}}
\newcommand\bea{\begin{eqnarray}}
\newcommand\eea{\end{eqnarray}}

\usepackage{epsfig}
\usepackage{gensymb}
\usepackage{footmisc}

\usepackage{color}
\definecolor{red}{rgb}{0.7,0,0}
\definecolor{blue}{rgb}{0,0,0.7}
\definecolor{purple}{rgb}{0.7,0,0.7}

\def\refe#1{#1}
\def\refee#1{#1}

\begin{document}

\title{Looking for observational signatures \refe{of early binary black hole systems }}
\titlerunning{Looking for \refe{early BBH} observational signatures}
   
\author{Peggy Varniere
          \inst{1,2}   
          \and
          Fabien Casse \inst{1}       
          \and
          Fabrice Dodu \inst{1}
          }

   \institute{Universit\'e Paris Cit\'e, CNRS, Astroparticule et Cosmologie, F-75013 Paris, France\\
             \email{varniere@apc.in2p3.fr}
         \and 
         Universit\'e Paris-Saclay, Universit\'e Paris Cit\'e,CEA, CNRS, AIM, 91191, Gif-sur-Yvette, France
}

\date{Received  / Accepted }

\abstract 
{A lot of recent studies have focused on the observables associated with near merger binary black-holes (BBHs) \refe{embedded in a circumbinary disk (CBD) but we still} 
we lack knowledge of observables \refe{of  BBHs in their early stage. In that stage the separation between the two black holes is so large that both black holes could potentially retain their individual accretion disk existing before the creation of the BBH. For such early BBH systems, it is interesting to look for observables originating in those individual disks whose structure is likely to differ from mini-disks often observed in simulations of later stages of BBHs.}
}
{ In a companion paper we presented \refe{ a set of hydrodynamical simulations of an individual disk surrounding a primary black hole while being impacted by the presence of a secondary black-hole in an early BBH system, leading to the creation of three well-known characteristic features in the disk's structure}.  
Here we explore the imprints of these three features on the observables associated with the thermal emission of the \refe{pre-existing black hole} 
disk.
The aim is two-fold, first to see 
 which observables are best suited for detecting those \refe{early} systems and, secondly, what could be extrapolated
  about these systems from observations.}
 {We used general relativistic ray-tracing in order to produce synthetic observations of the thermal emission emitted by \refe{early BBHs with} different mass ratio and separations  in order to search for distinctive  observational features of early systems.}
{\refe{We  found that in the case of early BBH with pre-existing disk(s)  a necessary, although not unique, observational feature is the truncation of their  disk(s)}.}
 {Such observable could be used for automated search of potential BBHs and discriminate some existing candidates.}

\keywords{accretion, accretion disks-- black hole physics  }

\maketitle
\nolinenumbers

\section{Introduction}

As existing and future gravitational wave (GW) detectors will provide rather uncertain sky localisations for the earliest detections of the sources,  it is therefore interesting to know 
what {would be the radiative emission} of a pre-merger binary black hole (BBH) system as it would help finding it faster within the GW localisation box through scan with electromagnetic telescopes.
At the same time, {such knowledge}  will give us access to {the structure and dynamics of}  the gas from the early to the last stage of the BBH merger.
As we want to be able to explore the impact of this strong, rapidly evolving, gravity on the accretion/ejection structures the best system to follow would be a BBH of $\sim 40 M_\odot$ 
 as they will go through a large dynamical range within the months prior to the merger{, allowing us to explore the transition from 
 \refe{early in the binary formation} to the last stages of the merger}. 
 The problem is that those systems might not be in a gas rich environment and are therefore not propitious for electromagnetic detection.
  Hence, if we want to study the dynamical evolution of the accretion-ejection structure at different stages of the pre-merger it is {therefore} more interesting to turn toward more massive 
  systems and 
  search  for a sample of {these} systems {being at} different stages rather than following one system through {its entire life span}.
In that respect we need to not only find what are the relevant observables associated with  {each}  stage but also check which ones will be easier to detect through a full scan of the sky or a targeted search. 
This paper is part of a series, each focusing on different stages\footnote{Refer to \cite{MVC24a,VMC24}  for observables related to the near merger stage \refe{when the BBH is embedded in a circumbinary disk}.} of the pre-merger system 
and looking for the optimal detection technic and period for those systems.

\refe{Up to now, the focus was on late}  
pre-merger systems\refe{, embedded in} a circumbinary disk\refe{, and their observables} \citep{shi_three-dimensional_2012,noble_circumbinary_2012,dorazio_accretion_2013,farris_binary_2014,noble_mass-ratio_2021-1,MVC24a}. 
\refe{As a lot of those observables are related to features of the CBD \citep{shi_three-dimensional_2012,MVC24a} and are not intrinsic to the BBH, it seems important to
look for the observational signature of BBHs in the absence of a CBD.
Indeed,  in the early phase of a galaxy merger, both black holes could feel each others' pull before gas could have the time to circularize around such a  wide binary, hence the early stage.
 Another case could be that the CBD actually exists but is too faint to be detectable due to the lack of gas in the vicinity, hence its observable cannot be used to identify the BBH.
 At the same time,  the individual black holes might still have retained their pre-existing disks and the impact of the other BH on these disks could lead to a new venue of detection for early BBHs.} 

\begin{figure*}[htbp]
\centering
\includegraphics[width=\textwidth]{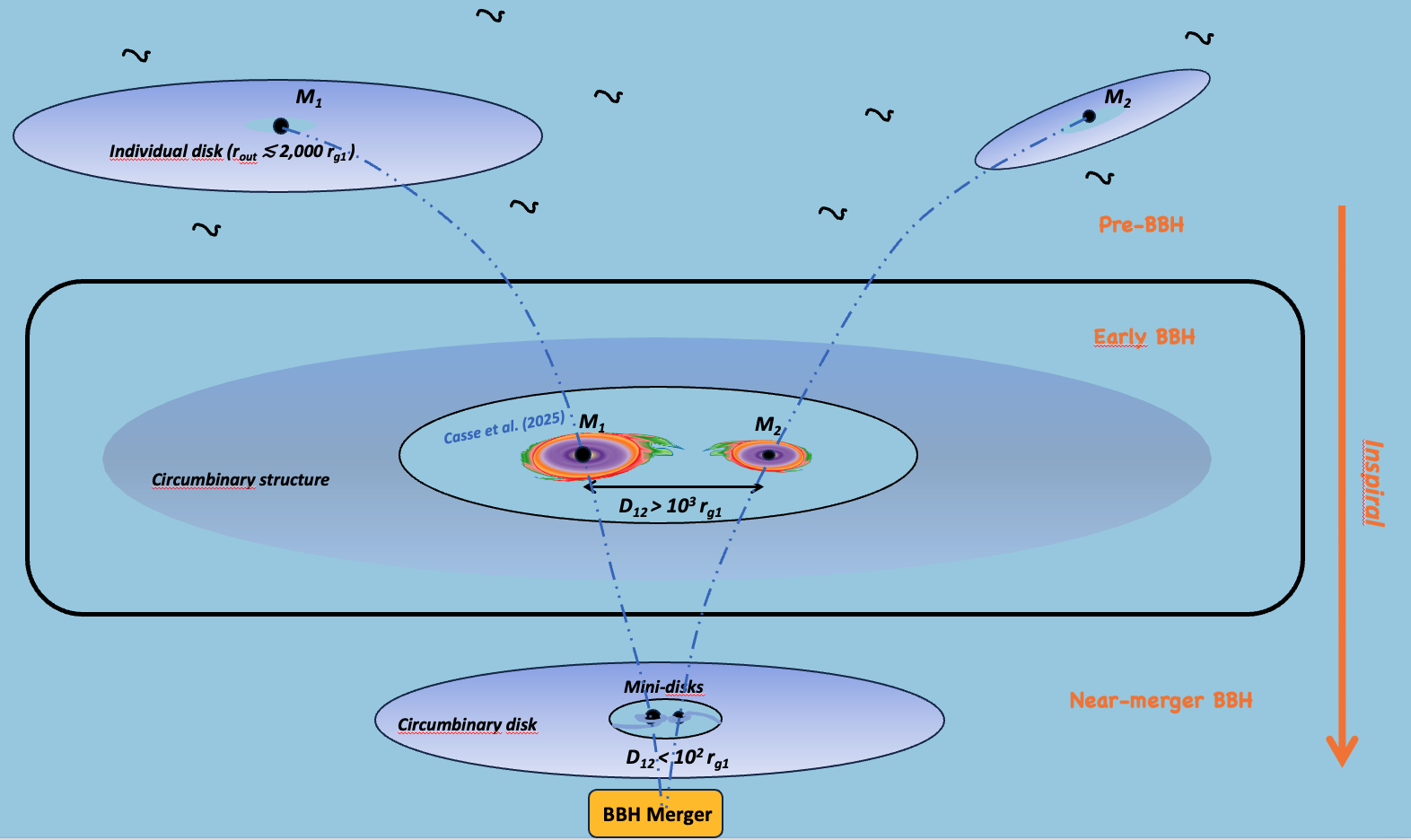} 
\caption{\refe{Schematic view of the main stages of a binary black hole system with masses of black holes are denoted by $M_1$ and $M_2$. In this paper we focus our attention on a quite unexplored stage of the BBH dubbed here as early BBH. In that stage that follows the birth of the BBH, the two black holes are gravitationally bound but still separated by a significant distance, namely $D_{12}\gg 10^3 r_{\rm g1}$ where $r_{\rm g1}=GM _1/c^2$ is the gravitational radius of the most massive black hole of the BBH ($G$ is the gravitational constant and $c$ the speed of light in vacuum). Such separation between the two massive compact objects allows each of the black holes to retain a significant part of their pre-existing disks if they had one. Based on hydrodynamical simulations of such disks performed in the context of an early BBH system \citep{CVAD24}, we explore the possible observational signatures originating from the presence of these disks subjected to the gravitational influence of both black holes.}}
\label{fig:BBHstages}
\end{figure*}

\refe{One of the reasons this early pre-merger time has not been as extensively studied as the CBD phase is that}  on one hand new observations of an early systems
 would not lead to a merger {occurring} over a human timescale for  supermassive BBHs and on a second hand stellar mass systems seem 
to be in a more rarefied gas environment, hence lacking electromagnetic emission.
Nevertheless, from those early systems stem the physical conditions from which circumbinary structures will be formed.   
Being able to detect such early systems would give us  {information on the earliest stage of the merger journey, especially how the accretion/ejection structures evolves along it.
Ultimately, this knowledge could be used to improve}
 the near-merger 
simulations that then could be used to identify systems closer to merger. 
{Having access to the observables of the early stage would also be helpful to do archival \refe{searches} ahead of the LISA mission, as a lot of the objects that would merge within
the LISA mission have spent some time in the early stage during the timespan of existing archival data.}
In the context of stellar mass BBHs, it is noteworthy that such detection of an early system with a gas-rich environment system\refe{, as unlikely as it is,} would be an invaluable asset as we would  be able to follow the system up to the merger {that would occur between 10 to 50 years once detected} hence providing access to its electromagnetic emission prior to the merger which is a pathway to the formation of the
circumbinary structure. 

 In a companion paper, \cite{CVAD24}, we investigated the impact of the gravity of {one of the black \refe{holes} composing an early BBH}  upon the accretion disk surrounding the other black hole. \refe{The simulations presented in \citet{CVAD24} were the first to depict the full structure of pre-existing disks in early BBHs from their last stable orbit to their outer edges for a large variety of black hole mass ratios and separations.}
In agreement with the earliest studies of 
 binary systems \citep{Paczynski77,Papaloizou77,Eggleton83,Sawada86,Spruit87,Whitehurst91,Artymowicz94,Lubow94,Godon98,Pichardo05,MirandaLai15,MartinLubow11,Ju16,Zhu16,Rafikov16,Bae18,Wester24}, 
\refe{we have identified  three main effects on pre-existing disks that are likely to alter their electromagnetic {thermal} emissions. 
With the help of analytical fits, provided by \citet{CVAD24}, quantifying the aforementioned features, we aim here to assess the detectability of 
these effects in order to see if they can contribute to the identification of early BBH systems.} Our paper is organised as followed:  
first we will briefly present the results from \cite{CVAD24} \refe{quantifying the impact} of the secondary disruptor upon the disk surrounding the primary black hole and \refe{displaying} on which timescales they occur.
In the following sections we will present the observational consequences of those three effects and assess their potential for identifying \refe{early 
BBH systems with pre-existing disk(s)}. 
\refe{Section \ref{sec:necessary} will discuss} a necessary observable of those systems and what could be \refe{learned} from such detection, in particular focusing on the outer disk detection in NGC $4395$ that calls for further
investigation. \refe{In the last section we briefly explore the impact of the presence of early BBH components, namely both potential pre-existing disks and CBD, on the aforementioned observables.}

\section{Impact of a secondary black hole on the circumprimary disk}
\label{sec:casse25}

 \refe{For the rest of the paper we focus on BBH systems with large separation, namely  larger than $10^3 r_{\rm g1}$  (see definition of the gravitational radius of the most massive black hole of the BBH system $r_{\rm g1}$  in Fig.\ref{fig:BBHstages}). We deemed those separations large in comparison with the typical size of accretion disks in active galactic nuclei (AGN)
 which tend to not have outer radii  larger than a few thousands of gravitational radii  \citep{Jha21}. }  
 
\noindent\refe{In this paper we define the \lq early stage\rq\   as the transitional period before a circumbinary disk is fully formed and interacting with the binary.}
 \refe{In early BBH systems one may expect a large cavity of about two times the separation \citep{Artymowicz94} to exist allowing the survival of a part of any pre-existing disks around each of the black holes.}   
 \refe{Fig.\ref{fig:BBHstages} presents a schematic view of the evolution of two black holes, with their pre-existing disks, starting from their pre-binary state then going to the early BBH stage we are interested in and finally to the near-merger stage, the most studied phase of  BBH where black holes rapidly inspiral toward each other while being embedded in a CBD.}
 
\noindent In \refe{this early BBH stage}, each black hole keeps their \refe{pre-existing} 
disk, if they had any, \refe{as they inspiral}. 
In order to see what happens to those original disks, in particular how long they remain and if they have any distinguishable features, we \refe{use the results from previously published}
 hydrodynamical numerical simulations of the accretion disk surrounding the most massive black hole of the binary taking into account the gravitational influences of the two black holes
  upon that disk. \refe{\cite{CVAD24} provides analytical fits of  the outer disk extension as well as its eccentricity for a wide range of black hole mass ratios and separations. This enables us to assess the extension and 
  eccentricity of both pre-existing disks, if they exist, in an early BBH system. As an illustration, both pre-existing disks showed in the early BBH stage of Fig.\ref{fig:BBHstages} originate from snapshots of 
  hydrodynamical simulations  from \cite{CVAD24} 
   showing the overall disk structure under the influence of both black holes.}

\subsection{2D fluid simulations of the early phase}

In \cite{CVAD24} we explored, through numerical simulation, how would the \refe{pre-existing} disk around a black hole be affected by the presence of a coplanar secondary black hole in a bound circular orbit  with a period of:
\begin{equation}
P_{\rm BBH} = 2\pi\left(\frac{D^3_{12}}{GM_1(1+q)}\right)^{1/2}=\frac{2\pi r_{\rm g1}}{c}\left(\frac{D^3_{12}}{r^3_{\rm g1}(1+q)}\right)^{1/2} \ .
\label{eq:PBBH} 
\end{equation}
  with $r_{\rm g1}$ the gravitational radius and $M_1$ the mass of the primary black hole on which the simulation is centered, $D_{12}$ the distance between both black holes and $q=M_2/M_1$ the black hole mass ratio.  We considered in this study only 
  the case where the secondary object is less massive than the primary, so $q \leq 1$.
   
 As we were looking for the impact on the outer edge of the  disk, the fluid simulations were done {in the frame associated with the primary black hole where its gravity is modeled using a 
 pseudo-Newtonian potential (see Eq.6 in \citealt{CVAD24}) while the gravitational force of the remote secondary black hole is included using a classical Newton law. In the present paper, ray-tracing  is achieved in a full general relativistic framework as this was shown to be sufficient to study {reliable synthetic observations of the disk} far from the black hole \citep{CV17}. Such choice prevented us from assessing the impact 
 of both the secondary black hole 
 and the spin of the primary black hole upon the inner region of the \refe{primary's disk} but enabled us to consider the dynamics of the outer disk over a large number of periods of the binary system. 
 Those approximations will be relaxed in a forthcoming paper fully devoted to the inner region of the  \refe{primary's disk}  
 whose observables will likely be detectable in much higher energy bands than  the effects presented here.
Similarly, we kept the secondary black hole {on a circular orbit in the same plane than} the 
 \refe{primary's disk} in order to look at the minimal effects, hence not taking into account potential warping of the outer edge which would amplify 
several of the effects presented here.

\subsection{Features of a circumprimary disk {in an early binary black hole system}}
\label{sec:resCasse25}

 \noindent   From the simulation done in \cite{CVAD24} we found that 
 the resulting outer edge of the disk is impacted by the presence of the secondary and takes {a typical shape as} shown in the middle panel Fig. \ref{fig:BBHstages}.
This leads us to three qualitative effects strongly linked with the presence, and parameters, of the secondary disruptor. 
\begin{description}
    \item[\tt -] First, we showed in \cite{CVAD24} (see e.g. {their} Fig.2) that not only the presence of 
 a secondary disruptor will shave off the outer edge of the \refe{pre-existing} circumprimary 
 disk, but we can also link the position of the outer edge with the mass 
 and position of that secondary object through a relation well fitted {for mass \refe{ratios} between $10^{-3}\leq q \leq 1$ by
$ r_{\rm out} = f(q)\ {D_{12}}$,
with $f(q)$\footnote{$f(q)= -0.18q^{0.044}\log{\left(0.019q+6.8\times 10^{-6}\right)} $ \label{footnote:fdeq}} a decreasing function of the mass ratio $q$.}

  While this is not the only mechanism
 that can truncate an outer disk, the presence of a secondary object close
 enough will always shave off the outer disk. 
     \item[\tt -] Secondly, after circularisation of the binary, the resulting smaller disk is not axisymmetrical but \refe{presents} an elliptical
     distortion that follows the secondary disruptor. 
     We showed in Fig.7 of \cite{CVAD24} that this eccentricity
     is quite high, at $e\simeq 0.6$, for {a secondary in circular orbit with a} mass ratio above $q=2 \times 10^{-2}$ and decreases for lower mass ratio.   
     \item[\tt -] Lastly, an $m=2$ spiral is present throught most of the simulation but with a small amplitude which makes it hard to see in the last few frames of Fig.2 from \citet{CVAD24}.
\end{description}
While those three effects have different amplitudes and timescales, they are all related to the orbital parameters  of the secondary and, if detectable, could lead to the 
{identification}  of \refe{systems in the early BBH stage.} 
 \\
\noindent \refe{The simulations presented by \citet{CVAD24} only considered the pre-existing disk around the most massive black hole of the system, but, by symmetry, the same results apply to the pre-existing disk of the second black hole if it exists.
}

\subsection{Raytracing with Gyoto}
\label{sec:gyoto}

{ In order to explore the impact of the three aforementioned effects on the observables of the system, we first need to compute the  spectral energy distribution (SED) of the resulting truncated disk.
This will be done with  the  general relativistic (GR) ray-tracing code {\tt gyoto} \citep{Vin11}  which we previously used in conjunction with the same pseudo-newtonian hydrodynamical code used in
 \cite{CVAD24}.
As it was shown in \cite{CV17}, it is not necessary to use full GR fluid dynamics when looking at the disk far away from a black hole, but having GR raytracing is still needed in order to fully capture 
what a distant observer would see.  As the fluid simulation \refe{was} done in the Paczynsky \& Wiita pseudo-newtonian approximation \citep{paczynsky_thick_1980}, hence with no spin for the black hole,
we compute the null geodesic in the Schwarzschild metric.}\\

\noindent {As we are  interested in  the  thermal  emission of the  truncated  \refe{pre-existing} circumprimary disk, we used the already implemented blackbody disk emission module from {\tt gyoto} in conjunction
with the disk characteristics found in the fluid simulations.
This reproduces a simple $\alpha$-disk model \citep{SS73} with a power-law temperature distribution\refe{:}
\refe{
\be
T(r) \simeq 6.3\ 10^5 \text{K}\ \left(\frac{\dot{M}}{\dot{M}_E} \right)^{1/4}   \left(\frac{M}{10^8\ M_\odot} \right)^{-1/4}  \left(\frac{r}{2r_{\rm g1}} \right)^{3/4}  
\label{eq:T}
\ee
with $\dot{M}_E$ the Eddington accretion rate. 
It is also fully compatible  with the {\tt diskbb}\footnote{\url{https://heasarc.gsfc.nasa.gov/xanadu/xspec/manual/node165.html}} 
model often used to fit observational data.}

\section{Evolution of the spectral energy distribution of \refe{one BH and its pre-existing disk} }
    
\noindent The first, and strongest,  effect we have seen in the simulations
is the truncation of the circumprimary outer disk linked with the position of the secondary black hole along the inspiral motion of the binary system\footnote{\refe{By symmetry, 
any pre-existing disk around the secondary will be truncated by the primary following the same principle (see sec.\ref{sec:resCasse25}).}}.
It is therefore interesting to assess the observational consequences of having a shrinking outer edge of the \refe{pre-existing} circumprimary disk during the {early} merger stage.
{This is especially interesting as the outer edge of the circumprimary disk is directly linked with the binary parameters and the time to merger}\refe{\footref{footnote:fdeq}}.

\begin{figure}
    \centering
    \includegraphics[width=0.47\textwidth]{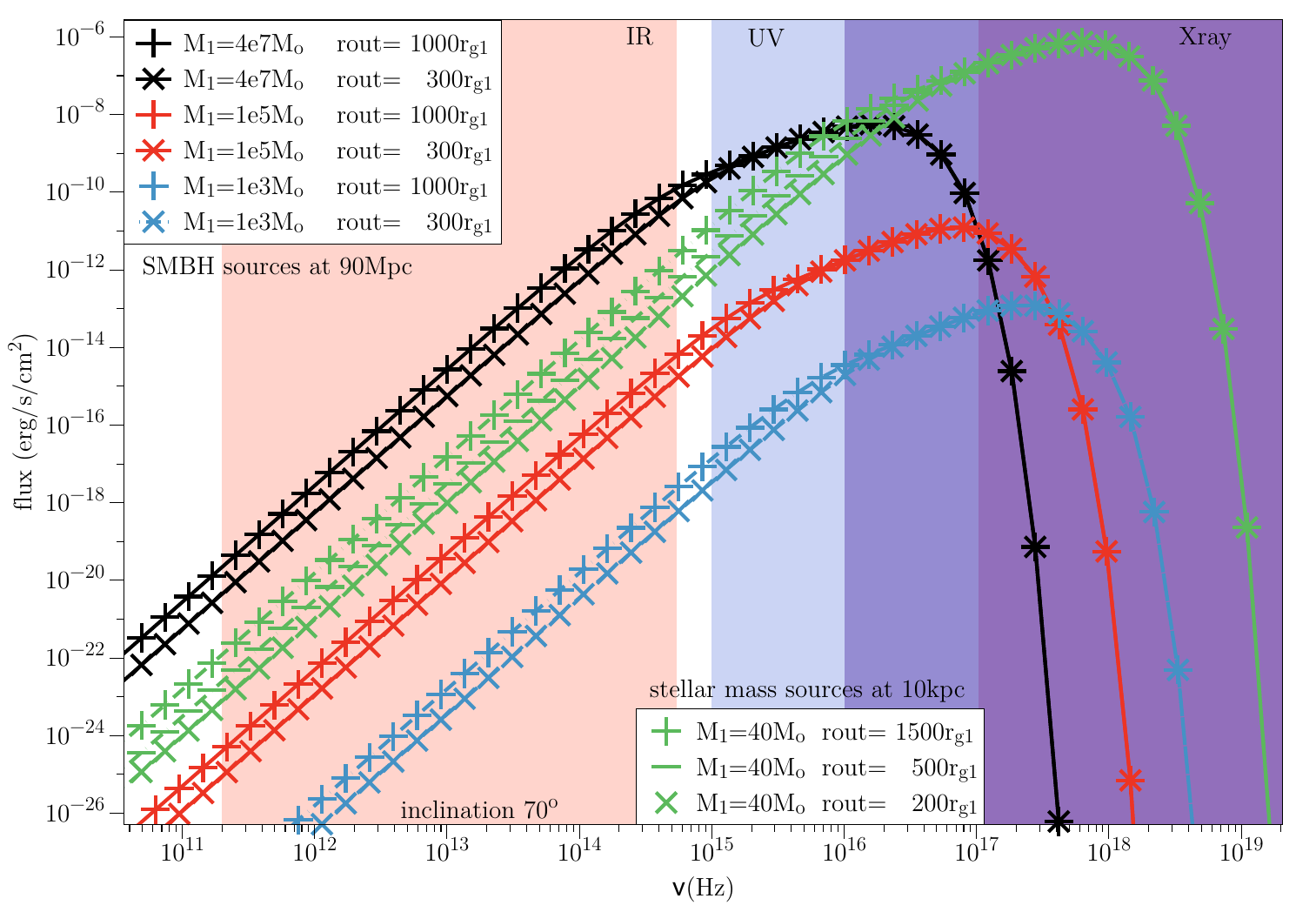}
    \caption{
        Evolution of the SED as the outer edge of the \refe{pre-exisitng} disk gets sculpted away for different masses of supermassive BBHs and radius evolution in the equal-mass case. 
    All the supermassive sources are located at 90.3Mpc while the $40 M_\odot$ is at 10kpc. All of them are seen at an inclination of $70^\circ$ to be easily compared.  
    }
    \label{fig:spectrumchange}
\end{figure}

 \subsection{SED of a truncated disk}
 
 Using  the results of the hydrodynamic simulations presented in \cite{CVAD24}
we produced synthetic observations using {\tt gyoto}   
in order to explore how the
missing outer disk \refe{impacts} the overall spectral energy distribution (SED) of the \refe{pre-existing} circumprimary  disk
assuming standard thermal emission for the disk (see sec.\ref{sec:gyoto} for details). 
Such computations will also be useful in order to identify which energy band in the multi-color blackbody spectrum is the more auspicious for a detection in the {early} phase. 
In order to find those energy bands, we displayed in Fig. \ref{fig:spectrumchange} how  
 the SED changes as the 
outer disk gets shaved off for a range of primary 
masses between $10^3 M_\odot$ and $4\times 10^{7} M_\odot$ located at a distance of $90$Mpc as well as for a lower mass of $40  M_\odot$ at $10$kpc.
As expected, one can notice that removing some of the outer disk causes the overall SED to 
decrease at lower frequencies, the exact energy where the emission starts to 
drop depending on the temperature of the missing outer disk, hence depending  on the mass of the central object.
For supermassive BBHs of masses above $10^6 M_\odot$, the changes will be mostly visible
in infra-red, while only primary black holes with masses lower than $ 10^6 M_\odot$ will exhibit a SED modification
starting in the lower end of the UV band.  
This contrasts with the late phase of the merger where a circumbinary disk is present and where most of the changes in the SED are predicted at higher energy 
\citep{roedig_observational_2014,farris_characteristic_2015,tang_late_2018,dascoli_electromagnetic_2018,gutierrez_electromagnetic_2022,dorazio_observational_2023,krauth_self-lensing_2023,cocchiararo_electromagnetic_2024,franchini_emission_2024}. 
\label{sec:long}

 \subsection{Constraint from the time to merger} 

While the changes coming from the removal of the outer disk shown in Fig. \ref{fig:spectrumchange} {appear to} be large enough to be detectable, we need to take into account the timescale 
on which the outer radius is changing.  Such process is occurring on the same timescale than the inspiral of the binary which depends on the mass of the binary. 
{In order to identify which kind of BBH system could exhibit an observable SED alteration we use the results found in \citet{CVAD24} to link the time to merger to the position of the outer radius of 
the circumprimary disk.  Indeed, {in \citet{CVAD24} we considered early BBH systems whose separations are up to a few thousand of BBH gravitational radii which is consistent with a gravitational wave emission dominated inspiral motion.} As we aim to find which system could be followed through the alteration of its SED, Fig. \ref{fig:time2mergerLM} 
{focuses on about the 100 years\footnote{this corresponds to \refe{an} average range of archival data} prior to the merger  and shows what would be the outer radius} 
of the \refe{pre-existing} circumprimary disk for  a selection of equal mass systems with the black hole primary mass ranging from $40$ to $10^7 M_\odot$.}
In this plot the vertical dashed lines represent the outer disk radius for the system used as examples in Fig. \ref{fig:spectrumchange} while the horizontal lines show the $100$ years and $50$ years time to merger. Such arbitrary time limits are of the same order than historical data and stand as a time limit for our ability to detect the movement of the outer edge of the circumprimary disk. 
Let us note that we have also represented the $10$ year 
time limit which correspond to the more recent data.

Looking at Fig. \ref{fig:time2mergerLM}, we see that the higher mass systems, namely with primary masses above $10^6M_\odot$, are already expected to have a very truncated original disk  one century prior to their merger\refe{, and would probably already have a CBD which would change the SED of the full system (see Sec.\ref{sec:fullBBH})}.
Regarding such systems it seems quite unlikely that we will be able to observe any significant evolution of their outer disk.
 This is especially the case as \refe{not only} the older observations would not have the same precision as more recent infra-red data\refe{, but  we would also need to add the typical stochastic noise of AGN}.
Moreover it is worth noting that for those higher mass systems the black hole separation is close
to the {lower} limit we can explore with our simulations and some other effects, linked for example with the formation of a circumbinary accretion structure, could lead to a different set of observables  
not studied in \cite{CVAD24} and therefore not present in the SEDs \refe{computed} 
here.
Similarly, when the supermassive binary black hole separation is larger than shown on Fig. \ref{fig:time2mergerLM} the change of the outer edge of the disk over a century is negligible and therefore will 
not lead to any detection. 
\begin{figure}
    \centering
   \includegraphics[width=0.48\textwidth]{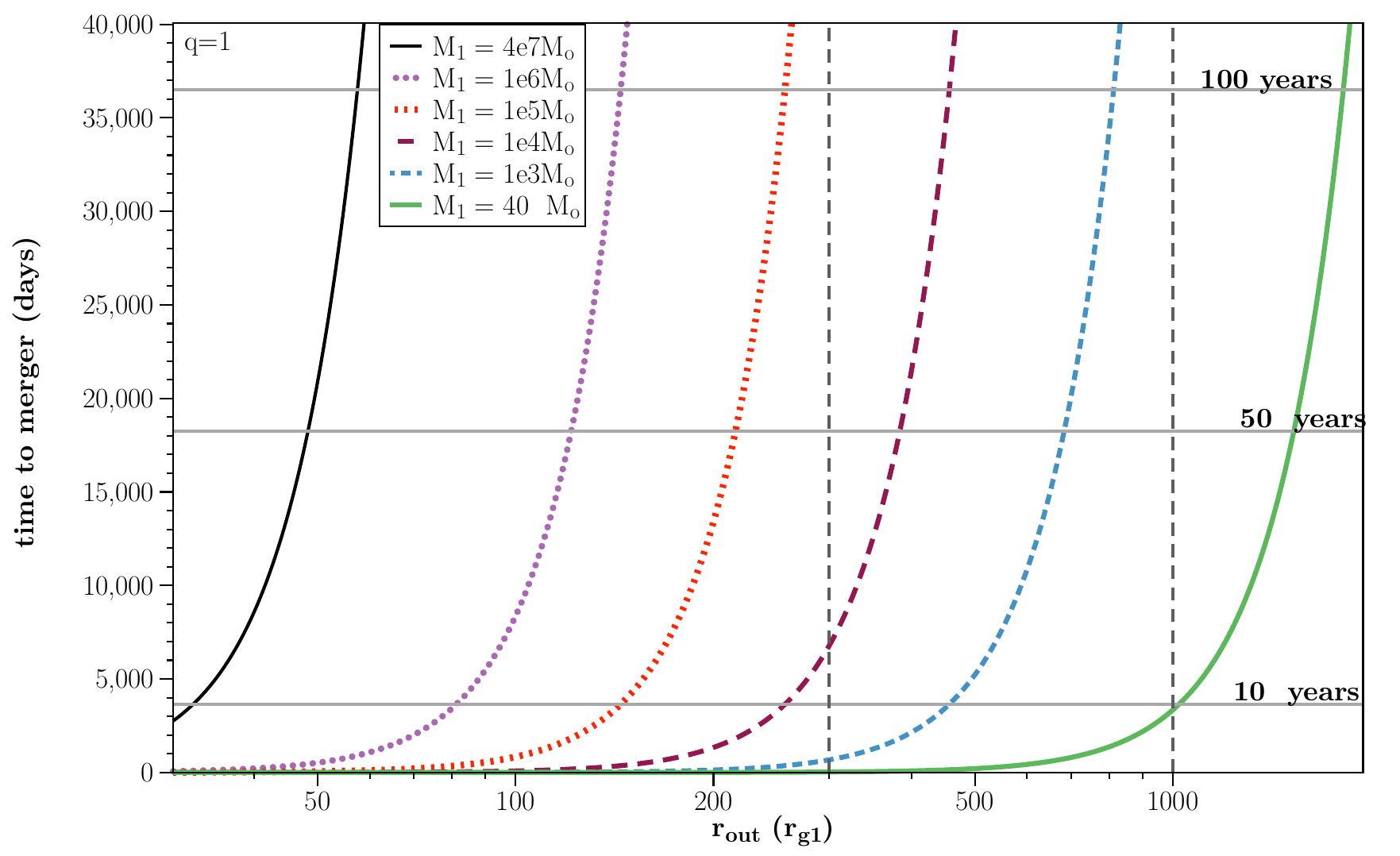}
    \caption{Evolution track linking the outer edge of the \refe{pre-existing} primary's disk and the time to merger for several equal-mass cases from low mass systems to the lower end of supermassive systems.
    }
    \label{fig:time2mergerLM} 
\end{figure}

On the other side of the mass spectrum, stellar mass BBH will exhibit a large enough change on a timescale of about $50$ years which could be looked for in historical data, especially as the shrinking of the orbit, hence the removal of the outer disk, gets faster as we get closer to the merger and some effect might even be detectable {within the last 10 years of astrophysical data}. 
Such a blind search through the optical and infra-red data archives {might} be automated. Indeed, even if a low mass BBH system with enough gas is 
not {likely} there is much to gain from the detection of the shaving of the outer disk in such systems.

 Similarly, intermediate mass systems 
would also be a good target for {such} blind search through the archive. 
Indeed, {despite evolving more slowly than {stellar}  mass systems, they are still expected to show} a significant variation of the outer edge of the circumprimary disk over 
a century. 
{As an example, a $2\times 10^3 M_\odot$ BBH system will go from  $r_{\rm out} \simeq 800\ r_{\rm g1}$ down to its merger over a century, which should lead to a potentially detectable change in the spectrum.}  \refe{Those systems could still have their \refe{pre-existing} disk or could have disks formed thanks to increased tidal disruption events as the binary separation decreases.}

 {As a conclusion and} despite the fact that the shrinking of the outer edge of the disks in BBH system is expected to be {a universal trend} in the early phase,  
{it appears difficult to follow its evolution for the most massive}  systems because of the timescale involved in such process.  
 The only systems where {such} timescale makes their detection conceivable are systems with potentially little gas such as stellar mass binary black hole, or 
  systems involving the elusive intermediate mass black holes.

\section{Impact at the BBH orbital timescale \refe{on the SED of the primary BH}}

As the timescale related to the shrinking of the orbit {makes it} too long to be detected early-on for supermassive BBH systems, 
we now {turn our attention on phenomena occurring on} a timescale related to the secondary orbital period (see Eq.\ref{eq:PBBH}).     
While it is still a long term, {namely} multi-years, timescale for the widest separations and largest masses (see Fig. \ref{fig:PBBH}), 
it is much shorter than the merger timescale {so that related phenomena} could be  {tracked} with carefully planned observations. 
On top of that, {the secondary orbital period becomes of the order of tens of days for a $10^7\ M_\odot$ BBH system at separation below a few hundred $r_{\rm g1}$ or below $4\times 10^3\ r_{\rm g1}$ for a $10^5\ M_\odot$ system, making associated periodic patterns potentially detectable for a wide range of binary parameters.}

\begin{figure}
    \centering
    \includegraphics[width=0.48\textwidth]{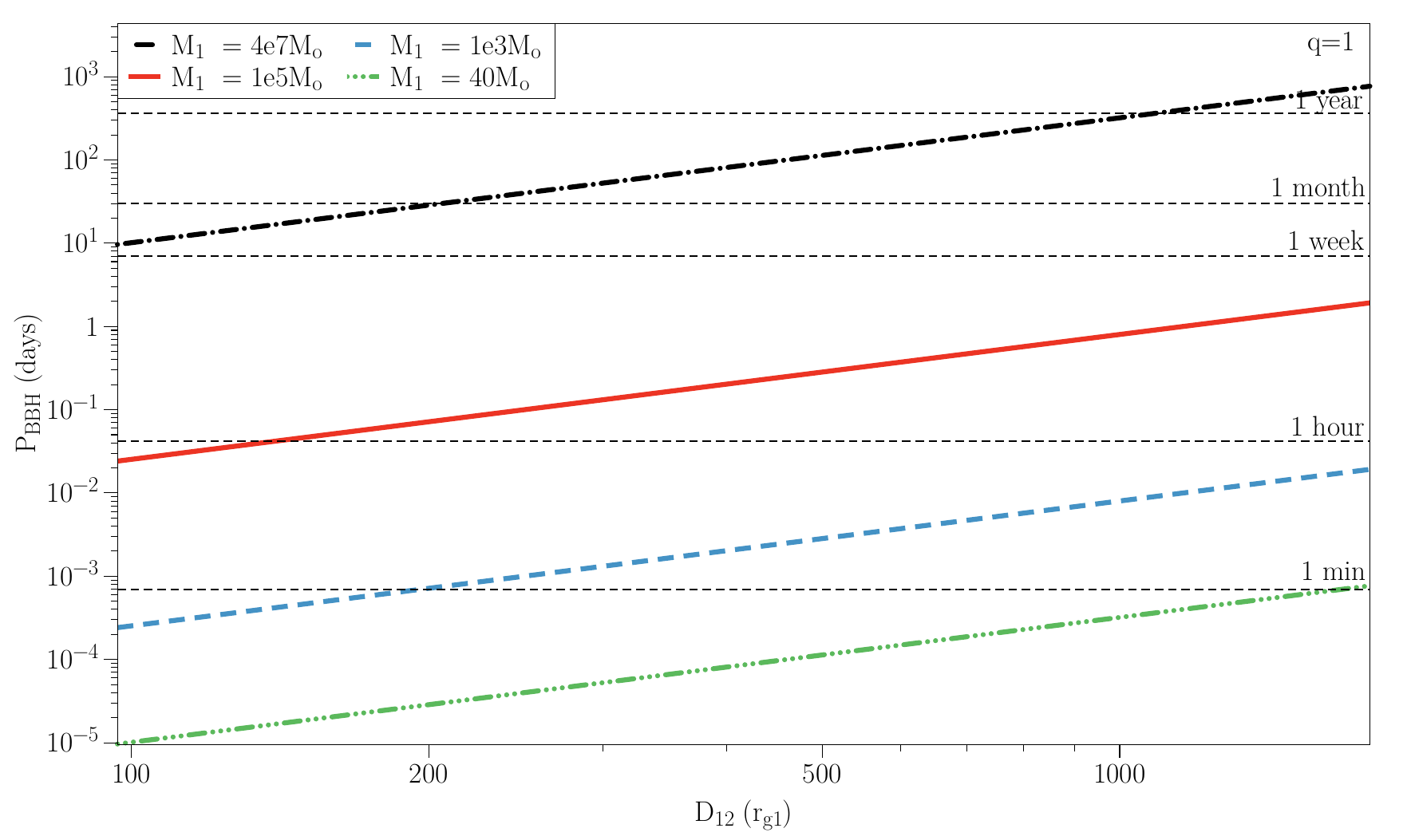}
    \caption{Evolution of the secondary BH orbital period as function of the binary separation in the case of equal mass binary. 
    While really short, less than one hour, for low mass systems,  it is within a year for a wide variety of masses and separations.}
    \label{fig:PBBH}
\end{figure}

\subsection{Impact of the presence of the spiral arms}

    Spiral arms are present in a wide variety of systems with different origins such as disk instabilities \citep{RWI_AEI} or the presence of a gravitational disruptor, bound or not to the system \citep{HD100546}
    and as such cannot be used as a final diagnostic of the {existence}  of a binary system. Nevertheless, they 
    will be present in case of a binary and, might be used to discriminate between systems {where a suspicion of binary is already present}.

     There are two major ways to look for spirals arms depending on how strong are those arms and how big
    is the system. The most common way for strong spiral structure is through its impact on the lightcurve
    as the spiral arm will create a modulation of the flux \citep{VV16}. This method 
    of detection does not work here as 
    the spiral arms are very faint and do not lead to any detectable-level modulation even for high inclination systems. 
    
    The second method is through direct detection of the spiral arm  
    which works well for massive companion, bound or not to the systems, as was shown in \cite{HD100546}.
    Even more recently there was a confirmation, in a young stellar object,  of the presence of spiral  
    arms driven by the companion \citep{YSO_apiral}.
    This method requires a large enough system as we need to directly observed the outer disk but also
    {requires the system to be suspected of being a binary system} as we need a long term observation campaign to follow at least part of the
    secondary orbital motion. 
    Both of those reasons make it unlikely that we will be able} to detect such effect for most early BBHs {as we not only need the system to
    already be a BBH candidate but also to be at a stage} 
    with favourable parameters.

 Therefore, the direct impact of the presence of the spiral arms {alone} does not seem to be a good diagnostic for the early-merger BBH.
    Another option, that we did not explore here, is to look for the impact of this spiral on the reflection spectrum, in particular the iron line profile. Indeed, it was shown in \cite{Karas2001}  that the non-axisymmetry of a spiral might lead to detectable features for future X-ray mission. This might be worth exploring in particular for high signal to noise systems as the spiral arms shown in our simulations would lead to 
      small and narrow features.

\subsection{A closer look at the impact of the ellipse precession}
\begin{figure}
    \centering
    \includegraphics[width=0.48\textwidth]{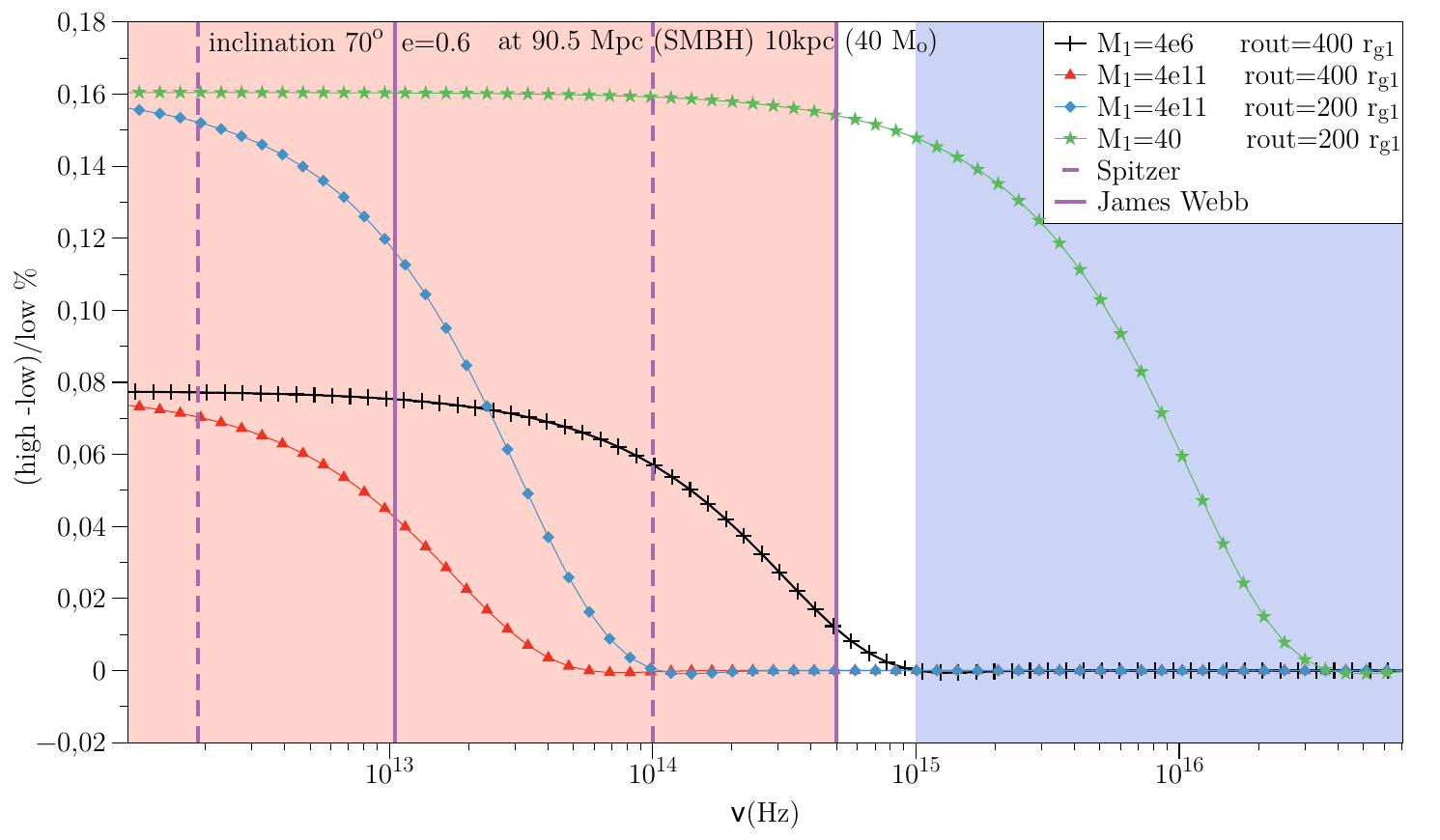}
    \caption{Relative difference between the \refe{pre-existing} {circumprimary disk} energy spectrum associated with the two extreme positions of its elliptical outer edge 
    with respect to the observer 
    {for various BBH masses}. This plot illustrates the impact of the ellipse precession on the energy spectrum in the case of equal-mass binaries. 
    We see a small effect over several order of magnitude in frequency, mostly below visible for supermassive BBH.
    }
    \label{fig:ellipse_spectra}
\end{figure}

  The second effect happening on the orbital timescale of the secondary disruptor is the
   precession of the elliptical outer edge of the disk once the binary is circularized. This is especially interesting as the {outer disk} eccentricity
   {induced by a second black hole in a circular orbit}
   can  reach $e \sim 0.6$ for $q> 10^{-2}$ as can be seen on Fig. 7 in \citet{CVAD24} and any non-axisymmetry present in the system 
   will be amplified by the GR boost as photons pass close to the black hole. Hence, even with the 
   low density seen in the elliptical part of the disk seen in Fig. \ref{fig:BBHstages} it is worth checking
   the level of the resulting modulation of the lightcurve.
   Especially as the change will be following the orbit of the second BH hence creating a periodic
   change in the shape of the energy spectrum that could help us determine the parameters of the secondary object. \\
  \noindent As the change in the {SED} happens over several orders of magnitude in frequency and flux
  it is easier to plot the contrast between the high and low phases of the modulation rather than the overall energy spectrum. 
  As a result, Fig. \ref{fig:ellipse_spectra} shows the maximum impact of the precession of the 
  outer disk on the shape of the energy spectrum. 
  As expected, the effect is small, with an amplitude of less than 1\% along the orbit of the secondary
  for an outer radius of the disk at $200\ r_{\rm g1}$. But it is worth noting that
  the amplitude of the effect increases with the shrinking of the outer edge
  hence the effect will be easier to detect as the system approaches 
  the merger. 
 
  For supermassive BBH the changes are expected  mostly in the infra-red with only the lower
  masses having some changes in the visible spectrum. 
  The case of {stellar} mass BBH will be more interesting with a large range of energy impacted from the
  very soft X-ray band to far infrared.  
  In that respect the band of the James Webb telescope will be able to reach a large number of systems
  from stellar masses to supermassive systems.
 
    Because of the low amplitude of the effect it will be difficult to use this variability in the shape of the energy spectrum as a diagnostic for the presence of an 
   {early} binary system.
   Nevertheless, as the effect grows stronger as the merger approaches it can be used as an additional
   test for BBH candidates {especially as other phenomena, such as {the ones generating  spiral arms}, will modulated the disk emission at the same frequency hence increasing the overall level of the detectable modulation}.
   In particular, if we focus on binary systems whose orbital period is up to a few tens of days, 
  we may have access to observations covering many orbits of the binary,  {which, in turn, would allow us to fold its lightcurve to ascertain} 
  the presence of such frequency with an improved signal-to-noise ratio.
   It is noteworthy that we {could also}  look for such periodicity with future wide-field or full-scan telescopes such as the future Roman telescope.

\section{The necessary observable of the missing outer disk}
\label{sec:necessary}
     When looking for a way to identify early BBH \refe{the first thought is to look for the changes in the the emission of the pre-existing disk of each black hole related to
     the inspiral of the binary.} 
      \refe{This leads to several problems related to
     both the long timescale on which one outer pre-existing disk varies and how small the effects related to the faster binary period are}\footnote{\refe{As a side note, the variability related to the potential CBD would be on a longer timescale than the binary period as the edge of the CBD is about 
     twice the binary separation \citep[see for example][]{noble_circumbinary_2012}.}}.
     \refe{While none of the effects presented here can be used as an unequivocal signature of a binary system, the} outer disk truncation \refe{represents} a {\bf necessary} 
     observable that could refute potential BBH candidate sources \refe{with individual accretion disk(s)} and 
     should be added as a check performed on all BBH candidates to see if their outer disk is indeed coherent with the inferred system size, mass ratio \refe{and time to merger}.

\subsection{A new way to look for early BBH candidates}

\subsubsection{{General procedure}}
 {In the quest for BBH systems, one could go a step further by exploring all systems exhibiting a smaller outer disk edge than average in order to see 
 what it would imply for the system to be an early BBH  
 and possibly triggering a search for other diagnostics}. \\

\noindent Indeed, knowing the position of the outer edge of the circumprimary disk, we can invert  the  
{relation between $r_{\rm out}$ and the binary parameters}
 to find 
  all {the mass ratio and binary separation in agreement with the reduced size of the disk}. Fig. \ref{fig:rout1700_whereBH2} 
  {illustrates, for four virtual observationally inferred outer edge radii ($r_{\rm out}=300, 500, 10^3, 1.7\times 10^3 r_{\rm g1}$), what would be the binary
parameter range $(q,D_{12})$ corresponding to each value. As an example, we have displayed in Fig. \ref{fig:rout1700_whereBH2} a shaded area encompassing the separation parameter  range corresponding to a circumprimary  disk exhibiting an outer edge where $r_{\rm out}=1.7\times 10^3 r_{\rm g1}$. According to our results the secondary black hole, for this case, would  be located at a distance between about $3\times 10^3$ and $5.5\times 10^3\ r_{\rm g1}$ depending on the mass ratio of the BBH system.}
While the mass of the secondary black hole is {impossible} to constrain this way, {its separation with the primary is limited to approximately 
two to three times the outer edge of the disk}.
\begin{figure}
    \centering
    \includegraphics[width=0.48\textwidth]{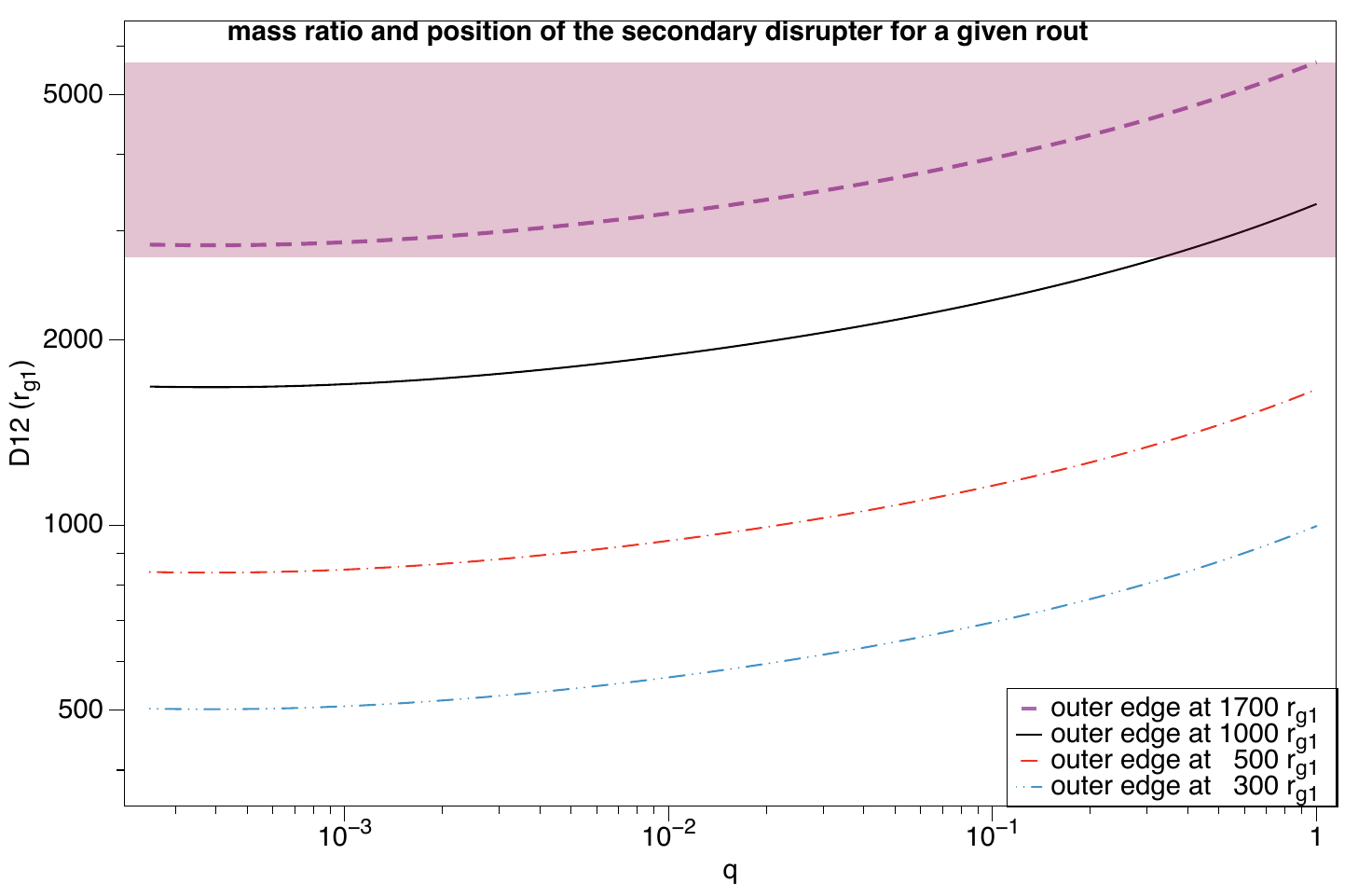}
    \caption{Possible parameter space for the secondary disruptor for a given outer edge $r_{\rm out}$.  We   show the line along which the secondary  should be for
     $r_{\rm out}=(300, 500, 10^3, 1.7\times 10^3)\ r_{\rm g1}$. If we have no prior knowledge of either the separation or the mass ratio, we obtain the shaded rectangle as potential
     zone \refe{for the secondary at the origin of the shrinking of the outer edge}. }
    \label{fig:rout1700_whereBH2}
\end{figure}

 {It is noteworthy that having the location for  the potential secondary black hole we may start looking for its presence through its impact on the circumprimary disk with a particular emphasis on variability as the BBH timescale, $P_{\rm BBH}$ (see Eq.\ref{eq:PBBH}), depends a lot on
 both the mass and separation of the binary system. Thus identifying such variability together with the presence of a reduced 
 \refe{AGN} disk point toward the presence of a secondary object whose mass may be estimated through the value of the variability period.} 
 Even {the non-detection of such variability} could put extra
 limits on the allowed parameters space for the potential binary {which in turn could either reject this possibility as a source of the truncation or help find a potential candidate.}

\subsubsection{The case of NGC 4395}

    Recently  \cite{NGC4395rout}  presented the \lq first detection of the outer edge of an AGN accretion disk\rq\  in the 
    case of the low-mass AGN NGC 4395.  Using results from HiPERCAM, the best reprocessing model gave a truncated disk with an outer edge around $1.7\times 10^3 r_{g}$
    with an error of $\sim 100\ r_{g}$. 
    While there is no hint to the presence of a secondary disruptor near NGC 4395, the existence of an outer edge detection for an  AGN  shows 
    that it will be possible in the future to test supermassive BBH candidates similarly. 
    For that reason, we are exploring here what would be the binary parameters if this outer edge is indeed caused by the presence of a secondary 
    disruptor.
    The shaded area of Fig. \ref{fig:rout1700_whereBH2}  shows the parameter space of  a secondary that would lead to a $1.7\times 10^3 r_{g1}$ truncation 
     for the circumprimary disk\refe{, namely a binary separation between $\sim 3000 - 6000 r_{g1}$ for a wide range of mass ratios}.
    While this possibility was not explored in their original paper, it would be interesting to see if any \refe{object is located at those distances, or if a} hint of the orbital frequency
    could be detected in the observables. This is especially interesting as NGC 4395 is a low-mass AGN, hence the orbital frequency of the secondary
    is of the orders of days\footnote{For a mass of NGC 4395 around $10^4 M_\odot$ we expect the orbital frequency of the secondary to be less than a day. 
    In case of a mass of r $10^5 M_\odot$ the period would be between 5 and 12 days.} at most which is detectable.

\subsection{From the elusive intermediate mass to the stellar mass black hole binaries}

\begin{figure}
    \centering
   \includegraphics[width=0.48\textwidth]{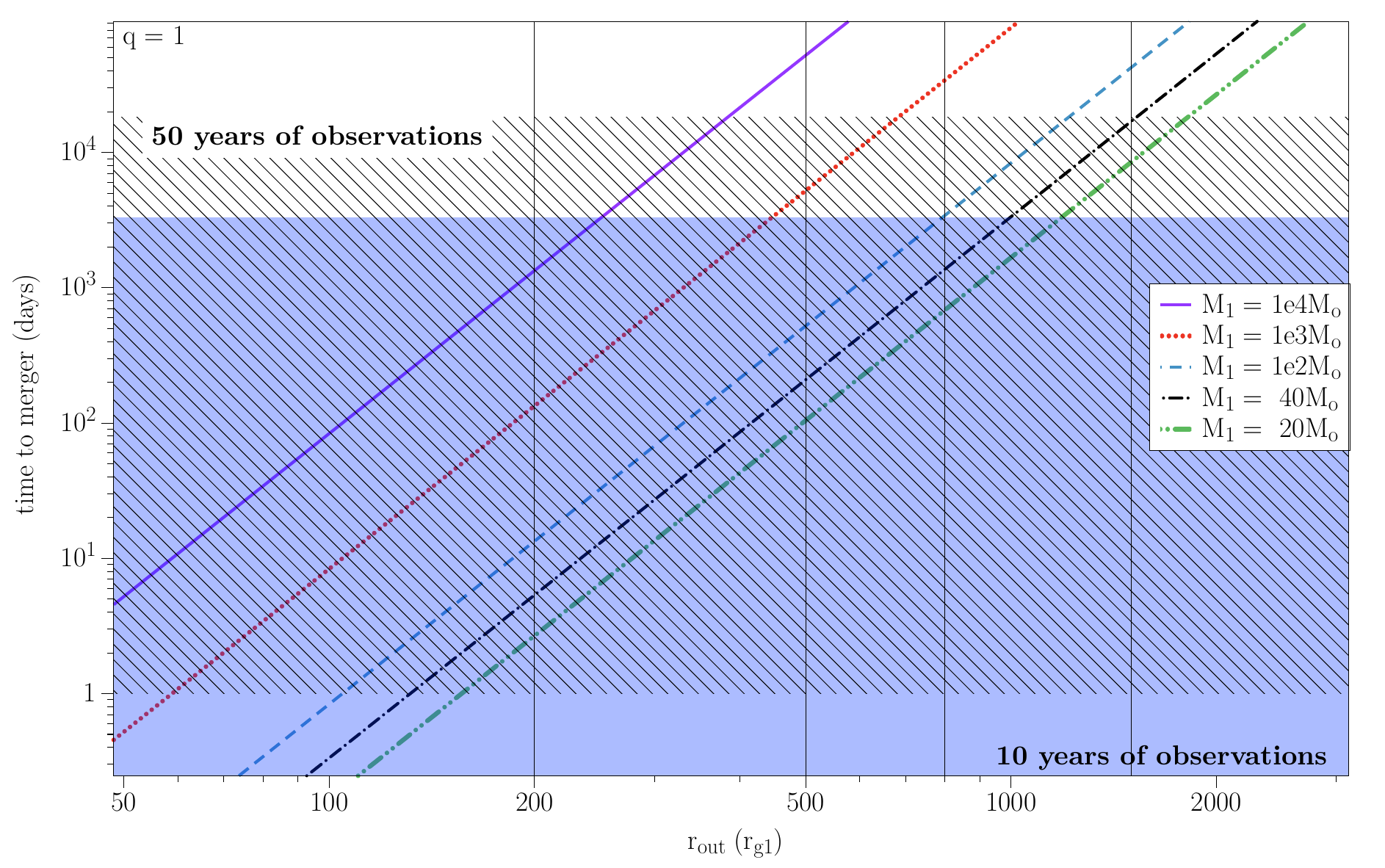}
    \caption{
    evolution of the time to merger as function of $r_{out}$  of the circumprimary disk for low equal-mass binary black holes.  
    The shaded area represent 10 years of observation prior to merger and the hatched one represent 50 years to merger (minus one day for visual clarity).
    }
    \label{fig:lowmass}
\end{figure}

While even 10 years ago we were only hoping to detect gravitational waves from binary black hole merger, they have now become 
{one of the best} messengers with which to study black hole systems. 
Still, we do not have {electromagnetic} pre-merger {BBH} detections, in part because the timescale between detection and merger of the running gravitational wave observatory is too small to catch the source before merger.

Hence, we would need another trigger to find those pre-merger sources with enough time to be able to do multi-wavelength  follow-up. 
For that reason, it seems interesting to look at what would be the emission of \refe{early pre-merger} intermediate mass black hole binaries even if those systems
{do not have a firm detection yet.} 
{Indeed, even if they} might not have {enough} gas {to render the black hole \lq identifiable\rq\  as such, it might be enough to trigger a 
search for smaller than average disk}. 
{On top of that, we can expect  strong changes in the SED  from UV to infrared (see Fig. \ref{fig:spectrumchange}) following the  
 fast evolution of the  outer edge of the  circumprimary disk for intermediate, and lower, mass binary black holes
 that can be seen on  Fig. \ref{fig:lowmass}.}
{Such changes makes an interesting criteria}  
for blind 
automated search through the optical and infra-red archival data and a positive results would give us access to 
a wealth of data with observations spanning $10$ to $50$ years before the merger. 
\begin{figure}
    \centering
    \includegraphics[width=0.48\textwidth]{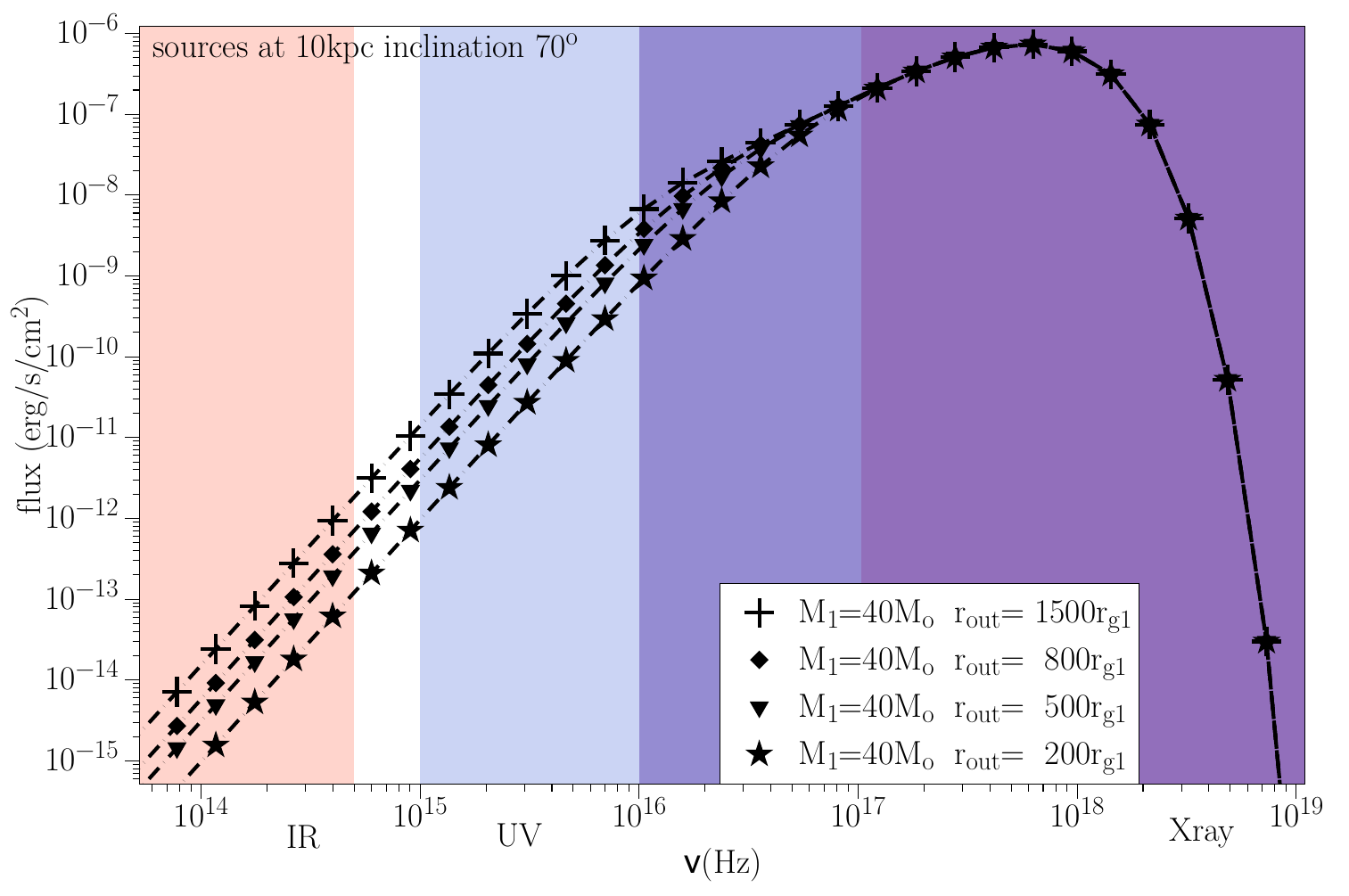}
    \caption{
     Impact of the decreasing $r_{\rm out}$ on the energy spectrum of the circumprimary disk in a system with a primary black hole of $40\ M_{\odot}$. {The time at which the spectrums are shown are represented on Fig. \ref{fig:lowmass} as perpendicular line, spanning other 50 years.}}
     \label{fig:stellarmass}
\end{figure}
  
\noindent  {The effect of the presence of a secondary black hole on the SED is even more striking on systems with the lowest mass \refe{if there is enough gas in the system 
for at least one of the black holes to have a disk\footnote{\refe{This could be, for example, soon after the supernova explosion forming the second black hole.}}.}
{Fig. \ref{fig:stellarmass} shows the overall expected change in the SED of a $40 M_\odot$ primary black hole}
 over about fifty years {prior to merger} (crosses versus star), {as well as some intermediary times (diamond at $\sim$ 3.5 years and inverted triangle $\sim$ 6 months to merger).} 
  While cross-correlating with older observations might be difficult, especially for  sources at great distance, it is worth noting that the 
   effect is accelerating as we get closer to the merger and the difference between {the diamonds and the stars} represent less than 4 years which is well within the life time of one instrument. 
   \refe{Once again, this is dependent on at least one of the black holes having a disk, and while this is not likely, even one detection would gives us a lot of information
   about pre-merger systems. }\\
      
   A search for such systems could also be done in preparation for the LISA mission which could detect the gravitational emission for the lower mass 
   BBH at a few months before the merger
    \citep{mangiagli_observing_2020,Mangiagli2022}. Because  of the large localisation box, having already identified \refe{potential} BBH candidates would help ensuring the best 
     \refe{pre-merger} electromagnetic follow-up \refe{if any gas is present in the system.}

 \section{\refe{Impact on the observables of the full binary system}}
 \label{sec:fullBBH}

\refe{Up to now we have considered how the different phenomena presented in sec.\ref{sec:resCasse25} impact the observable of the system comprised of one black hole and its
pre-existing disk. While this could represent a very early case where the angular separation between both black holes could allow for separate observations, it  was mostly presented to
showcase the individual effects.
Here we look at the more general case when we are not able to separate the components of the binary system.}

\begin{figure}
    \centering
   \includegraphics[width=0.49\textwidth]{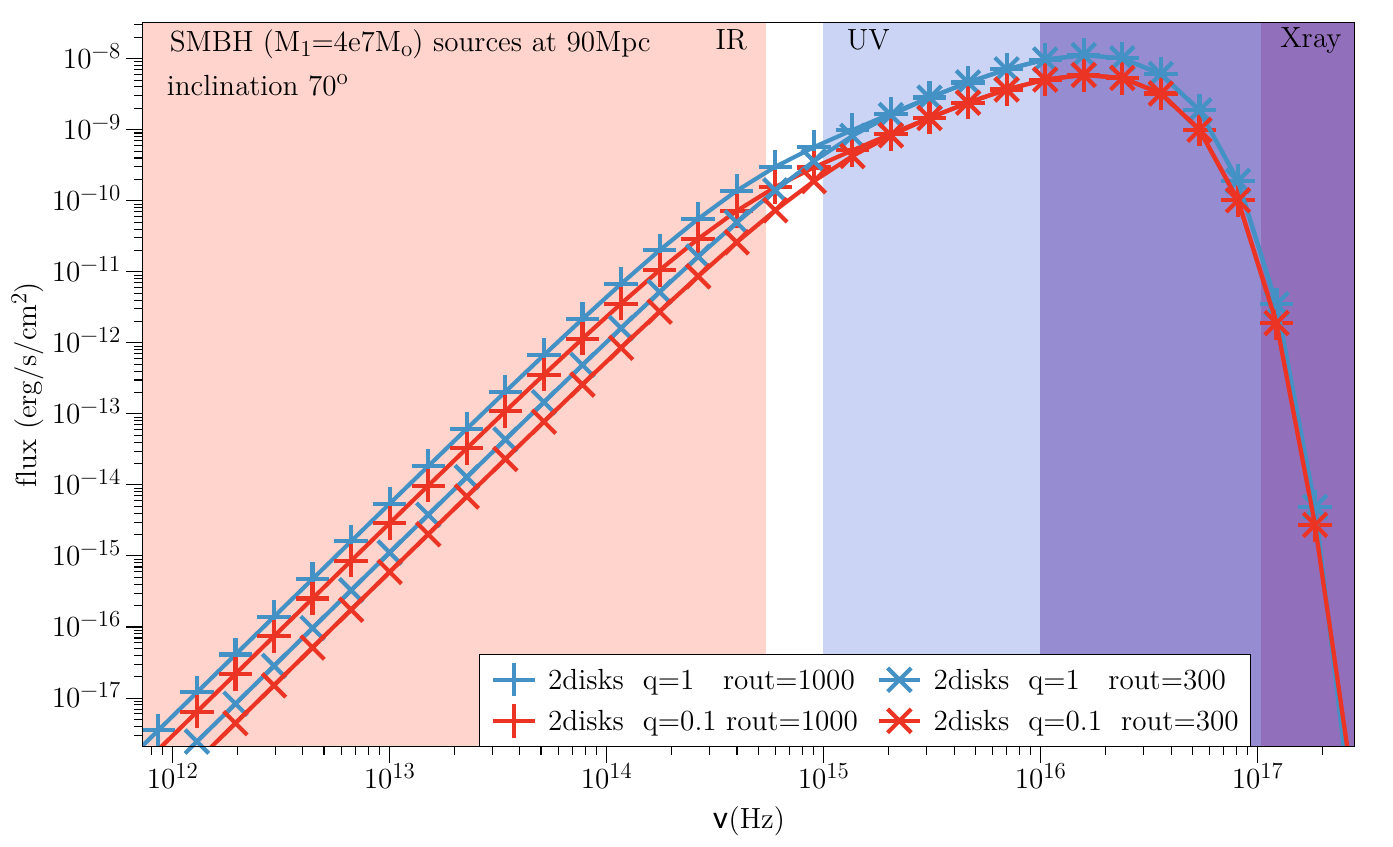}
    \caption{\refe{Evolution of the SED in case both black holes and their eroded pre-existing disks are visible in the same observation for 
    two mass ratios $q=1$ and $q=0.1$.} }
    \label{fig:2disksSED}
\end{figure}

\subsection{\refe{Impact of the presence of the secondary black hole in the observation}}

       \refe{First we look at the case where the secondary black hole also has a pre-existing disk but a circumbinary disk is not detectable, either because it is not fully
       formed yet or it is too faint.}
      \refe{All the results from \cite{CVAD24}  summarized in sec.\ref{sec:resCasse25} as well as the results presented above 
       are also valid for the disk of the secondary black hole.
       Hence, both of the outer edges will be shrunken as the binary inspirals closer together and the SED of both will exhibit a similar decrease in the lower energy band.}
       
 \noindent \refe{In order to showcase this effect, while making it easier to compare with the previous plot, we chose to keep the mass of the primary (heavier) 
 black hole at $4\times10^7 M_\odot$ as was used in Fig.\ref{fig:spectrumchange} and also at the same $r_{out}$ for the primary black hole. 
 Fig.\ref{fig:2disksSED} show the evolution of the SED for two binary black hole system, the  \lq extreme\rq\ case of equal mass case and a smaller, $q=0.1$,
       mass ratio.  The equal mass case is easy, as each black hole and its pre-existing disk are interchangeable, hence we have twice the same flux. 
      For the other mass ratio we first use the relation found in  \cite{CVAD24} between $r_{\rm out,1}, q$ and $D_{12}$\footref{footnote:fdeq} to obtain the separation between both 
      black holes, which we then translate into units of the  gravitational radius of the secondary black hole $r_{\rm g2} = GM_2/c^2= q r_{\rm g1}$. From there we have all we need to compute the
      position of the truncation of the disk of the secondary black hole. 
       For simplicity we assumed the disk of the secondary to be in a similar state with a mass-scaled temperature from the $\alpha$-disk model \citep{SS73} 
       and following Eq.\ref{eq:T}.
}
\newline

     \noindent \refe{As expected,   Fig.\ref{fig:2disksSED}  shows that, even if we cannot isolate one of the black holes, we will still be able to see the drop in 
       the lower energy band of the SED and it will even be greater than the one from Fig.\ref{fig:spectrumchange} in the equal mass case.  
       In fact we can say that the $q=1$ SED from  Fig.\ref{fig:2disksSED}  is the  \lq best case scenario\rq\  for  the detectability, when both effects add up at every energy, 
       while  Fig.\ref{fig:spectrumchange} represent the worst case scenario, where the secondary disk does not contribute significantly to the SED.
                 }

\subsection{\refe{Impact of the presence of the CBD in the observation}}

      \refe{While the focus of the paper is to search for observable signatures of BBHs having no, or at least not detectable, CBD we add, for completion, 
      how the presence of a CBD would  change the SED presented above for two black holes and their disks.
      First, the shrinking, being only related to the gravitational impact of the secondary,  would still exist when the binary is surrounded by a CBD, though
    the emission from the CBD would need to be taken into account. }
     \newline

    \noindent \refe{Once again, for ease of comparison, we choose an equal mass binary with the primary black hole at $4\ 10^7 M_\odot$  
      and we also keep the same evolution in $r_{out}$ as in the previous figures.  
      For simplicity we chose to have the CBD be in a similar state as the previous disks but orbiting around the center of mass of the binary at twice the
      binary separation and used  the same mass-scaled $\alpha$-disk model \citep{SS73} but for the total binary mass (see Eq.\ref{eq:T}).
      \refee{Depending on how much material was available in the environment around both black hole and disk, this simple approach could be an over, or under, estimation
      of the density in}    
      the CBD in the early BBH stage \refee{but it gives an order of magnitude for the SED.}
      }  \\
      x
     \noindent \refe{Nevertheless, not only 
      Fig.\ref{fig:CBD_2disksSED} shows that we recover the \lq notch\rq\ often presented as the diagnostic of the configuration of two inner disks and a CBD with a cavity in between, 
      but, on top of that, we see that, as a consequences of the inspiral and the 
      shrinking of the outer edge of both black hole disks: 
  \begin{description}
      \item[\tt 1)] the \lq notch\rq\ is moving toward higher energy
       \item[\tt2)]  the decrease of the SED due to the reduced size of the black hole disks is only visible  above the \lq notch\rq.
 \end{description}     
      While the effect of the shrinking outer disk is detectable in a more limited in energy range due to the presence of the CBD, it also gives us a stronger 
      observable of inspiraling BBHs when it is associated with this moving  \lq notch\rq. 
      As both movements, the inner edge of the CBD and the outer edge of each black hole disks, are on the same timescale, all the considerations from the previous 
      section are still valid when both effects are taken into account.  
     }

\begin{figure}
    \centering
   \includegraphics[width=0.49\textwidth]{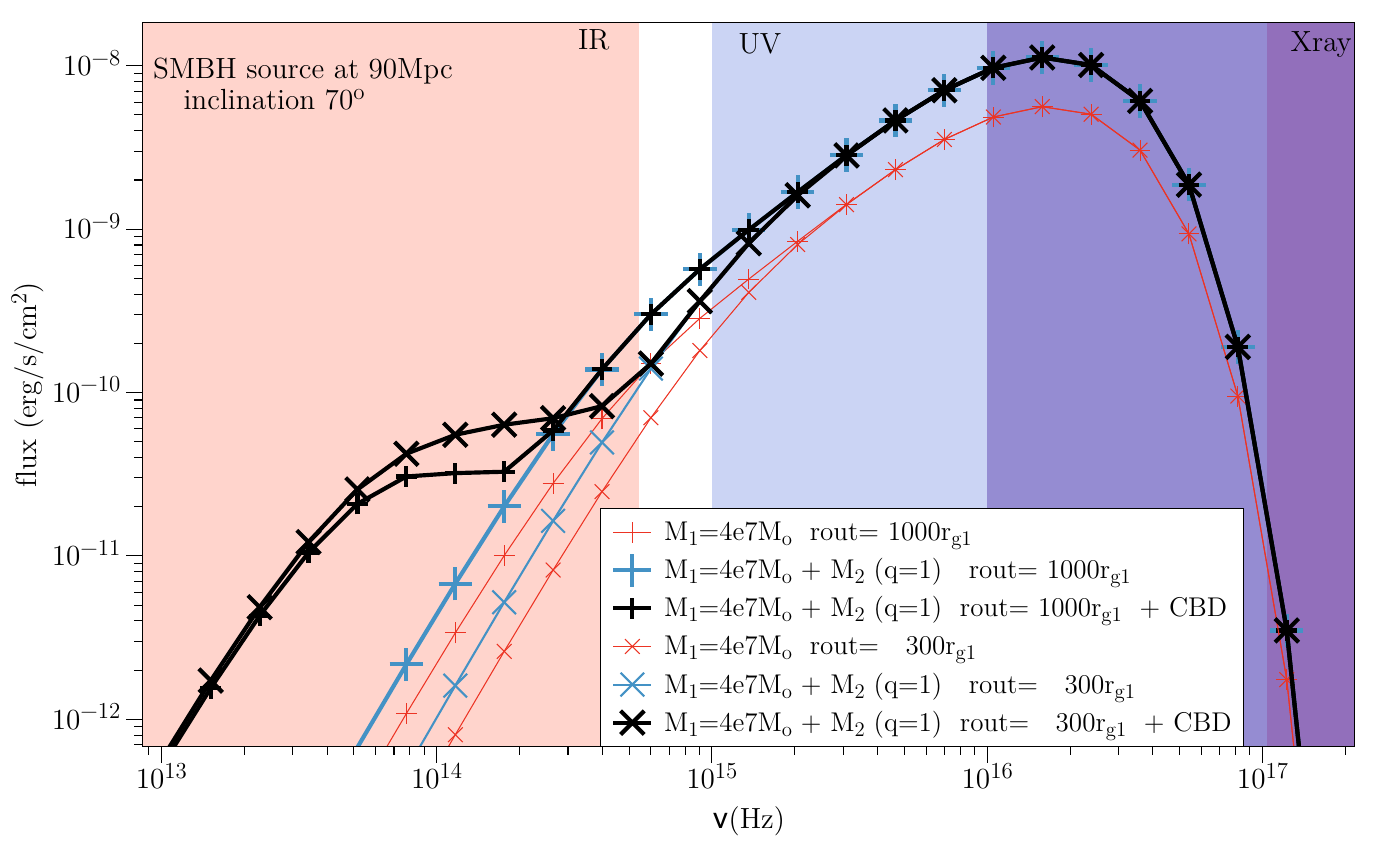}
    \caption{\refe{SED in case of equal mass black holes with each an original eroded disk and their CBD is detectable.}}
    \label{fig:CBD_2disksSED}
\end{figure}

\section{Conclusions}

With the first detection of gravitational wave from compact object merger 
\refee{the hunt for pre-merger BBHs received a lot of attention from the community as we finally had observational confirmation of their existence.}
In order to detect such systems, we \refe{first} need to characterise \refe{their} observables and, in particular, assess their unicity. 
\refe{Up until now the most promising observable that could help identify near merger BBHs is linked to the variability of their circumbinary disk.
Here we are interested in earlier systems which allow us to present a complementary observable linked this time with the pre-existing black hole disk 
and how they are affected by the presence of two black holes.}
\\
\noindent  In a companion paper, \cite{CVAD24}, we used numerical simulation to explore the impact of a gravitationally 
{bound secondary black hole on the original accretion disk surrounding the primary black hole}.
We \refe{not only confirmed the presence but also quantified}  three qualitative effects occurring  on two different timescales, all of them  strongly linked with the {characteristics} of that secondary black hole. \\

\refe{Using those previous results we compute} 
synthetic observations of \refe{early BBH} systems. \refe{We} looked {at  the impact of these three effects upon the thermal emission of \refe{a pre-existing 
circumprimary disk} and how {it} could lead to \refe{early} BBH detections.
We found that the spiral wave going through the {circumprimary} disk and the {eccentricity of} its  outer {edge}, both occurring on the secondary {black hole} orbital 
timescale,  have too weak of an effect to be a
significant signature of  early BBH. 
Nevertheless, both of those effects will become stronger as we get closer to the merger, so they might eventually become detectable and could be,
by their absence,  a way to give an upper limit estimate of how far from the merger the system is.

More interestingly, we found a {\bf necessary} observable for {early} \refe{(wide binary with a separation greater than $10^3 r_{g1}$)} BBH \refe{having pre-existing disk(s)}. Indeed, the presence of a \refe{near equal mass} companion 
\refe{($ q \geq 10^{-3}$, \citealt{CVAD24})}  will  slowly shave the \refe{pre-existing} 
{circumprimary outer disk} as they are getting closer to one another, hence making the missing outer disk a strong signature of  \refe{early BBH} system. 
While this phenomena is not unique to gravitationally bound systems as a fly-by of another massive object, 
would create similar, though transitory,  effect, it is nevertheless assured to exist in a binary system.  Hence, it should be used, when possible, as a test of binary for all  
\refe{early} BBH candidates to search for any incoherence between the inferred system parameters and the expected disk size. 

\noindent \refe{Once the binary reached a stage where the emission from the circumbinary disk is detectable, while also having a wide enough separation for each black hole to still
have their  \refe{pre-existing} disk, it becomes more interesting to focus the search for those systems on the changes of the SED around the \lq notch\rq\  area.
Indeed, not only we will still see the effect of the shrinking of the outer disk, though in a narrower energy band, but the \lq notch\rq\  will also be moving toward 
higher energy as the binary inspirals.}
\\

While finding the radial \refee{extent} of AGN disk is not an easy task, the fact that it could be used to flag \refe{early} BBH candidates among supermassive black hole systems makes it an interesting goal. 
This is especially true, if a method could be automated 
to search through archival data for  \lq smaller\rq\ than average or receding 
disks in order to flag potential \refe{early} BBH candidates which could \refee{then} be independently
targeted to prepare for future GW mission such as LISA.

\section*{Data Availability}

The data that support the findings of this study are available from the corresponding author, PV, upon request and will also be part of a data release in mid 2026\footnote{Which will be 
available for download at  {https://apc.u-paris.fr/$\sim$pvarni/eNOVAs/LCspec.html}}.

\begin{acknowledgements} 
The numerical simulations we have presented in this paper were produced on the DANTE platform (APC, France)
and on the high-performance computing resources from GENCI - IDRIS (grants A0150412463 and A0170412463).
Part of this study was supported by the LabEx UnivEarthS, ANR-10-LABX-0023 and ANR-18-IDEX-0001.
\end{acknowledgements}


\bibliography{Outer_Edge} 





\end{document}